\documentclass[9pt,twocolumn,twoside]{opticajnl}
\journal{opticajournal} 

\setboolean{shortarticle}{true}

\usepackage{endnotes}
\let\footnote=\endnote

\title{Relative Wavefront Error Correction Over a 2.4~km Free-Space Optical Link via Machine Learning}

\author[1,2]{Nathan K. Long}
\author[3]{Benjamin P. Dix-Matthews}
\author[3]{Alex Frost}
\author[3]{John Wallis}
\author[1]{Ziqing Wang}
\author[1]{Kenneth J. Grant}
\author[1,*]{Robert Malaney}

\affil[1]{School of Electrical Engineering and Telecommunications, University of New South Wales, Kensington, NSW, Australia}
\affil[2]{Defence Science and Technology Group, Edinburgh, SA, Australia}
\affil[3]{International Centre for Radio Astronomy Research, University of Western Australia, Perth, WA, Australia}

\affil[*]{r.malaney@unsw.edu.au}

\begin{abstract} 
In coherent optical communication across turbulent atmospheric channels, reference beacons can be multiplexed with information-encoded signals during transmission. In this case, it is commonly assumed that the wavefront distortion of the two is equivalent. In contrast to this assumption, we present experimental evidence of relative wavefront errors (WFEs) between polarization-multiplexed reference beacons and signals, after passing through a 2.4~km atmospheric link. We develop machine learning (ML)-based wavefront correction algorithms to compensate for observed WFEs, via phase retrieval, resulting in up to a $2/3$ reduction in the relative phase error variance. Further, we analyze the excess noise contributions from relative WFEs in the context of continuous-variable quantum key distribution (CV-QKD), where our findings suggest that if future CV-QKD implementations employ wavefront correction algorithms similar to those reported here, an order of magnitude increase in secure key rates may be forthcoming.
\end{abstract}

\setboolean{displaycopyright}{false}

\begin{document}

\maketitle


Free-space optical (FSO) communication across turbulent atmospheric channels often involves encoding information on the electric field quadratures of an optical signal (hereafter, ``the signal''), which can be polarization-multiplexed with a phase reference beacon (hereafter, ``the reference'') before transmission. It is commonly assumed that the distortion experienced by the spatial phase wavefront of the reference is equivalent to the distortion experienced by the spatial phase wavefront of the signal. As such, the reference wavefront measurements can be used to modulate a real local oscillator (RLO) at the receiver such that the signal electric field quadratures can be coherently measured using balanced homodyne/heterodyne detection.

Previous works have investigated relative WFEs between the reference and signal wavefronts due to differing reference and signal wavelengths, which resulted in differential interactions with atmospheric turbulence (e.g.~\cite{Han2024}). Beyond wavelength differences, various additional factors could lead to the reference and signal wavefronts experiencing differential distortion, regardless of wavelength, resulting in relative WFEs between them. For example, imperfections in optical hardware used at the transmitter and receiver, imperfect polarization multiplexing, photon leakage between the reference and weaker signal, and differential interactions with atmospheric turbulence (such as due to weak birefringence) could result in relative WFEs. 
\textcolor{black}{To explore this further, we conducted a series of experiments to investigate whether or not relative WFEs can be detected across a 2.4~km turbulent atmospheric link. As we show below, we answer this question in the affirmative. We subsequently characterized the relative WFEs and developed ML-based wavefront correction algorithms, which could be applied to an RLO for increased coherence with the signal during measurement. }

\begin{figure*}[htbp]
    \centering
    \includegraphics[scale=0.8]{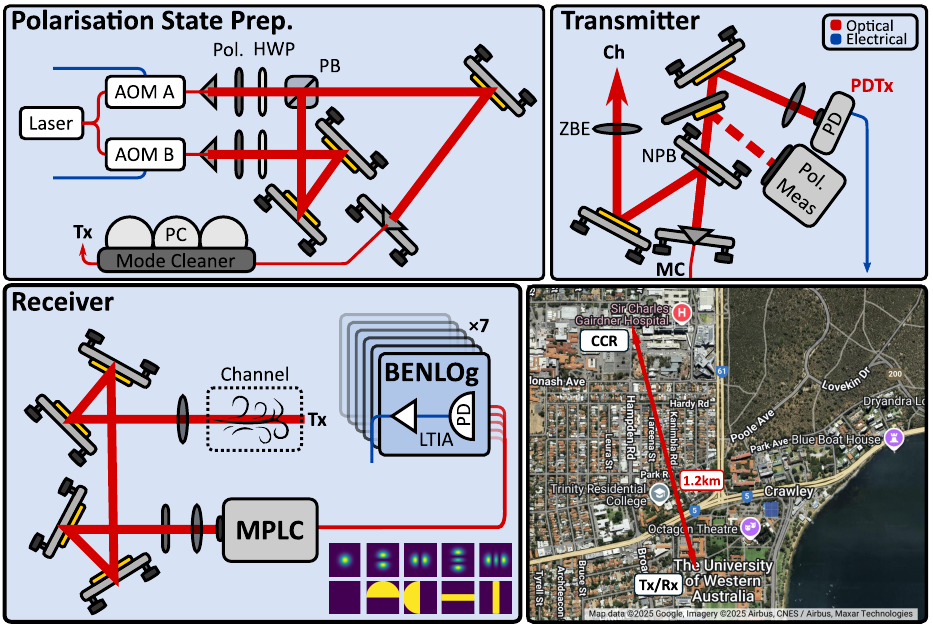}
    \caption{Experimental setup for preparation and measurement of a reference and signal, with the 2.4~km link shown at the bottom right, and HG mode intensity and phase profiles shown at the receiver. AOM is an acousto-optic modulator, Pol is a polarizer, PC is a polarization controller, PD is a photodetector, ZBE is zoom beam expander, MC is the mode cleaner, Tx is the transmitter, Rx is the receiver, Ch is the channel, CCR is the corner-cube retroreflector, and BENLOg represents our logarithmic photodetectors.}
    \label{fig:exp_model}
\end{figure*}

Adaptive optics (AO) is commonly used to measure reference wavefronts, followed by an imprint of the wavefront information onto an RLO, for example, using a deformable mirror. Shack-Hartmann wavefront sensors (SH-WFSs) are often used as wavefront sensors in AO. However, SH-WFSs are sensitive to non-uniform illumination~\cite{Akondi2019}, where scintillation often increases as channel length increases. Multi-plane light converters (MPLCs) are emerging as an alternative to SH-WFS in AO for FSO communications~\cite{Cho2022}. MPLCs can decompose an electric field into any arbitrary spatial mode basis, where we utilized an MPLC in this work to decompose the incoming electric field $E(x,y)$ into the Hermitian-Gaussian (HG) spatial mode basis using $N$ modes \textcolor{black}{(as in~\cite{Cho2022})}, ${E(x,y) = \sum^N_{mn} c_{mn} \mathrm{HG}_{mn}(x,y)}$, where $\mathrm{HG}_{mn}(x,y)$ are the transverse profiles of the HG mode, and $c_{mn}$ are complex weighting coefficients for each mode. 

We polarization-multiplexed an optical reference and a weaker optical signal (transmitted at approximately -3.1~dBm and -8.1~dBm, respectively --- see AOMs in Fig.~\ref{fig:exp_model}) during transmission, both prepared by the same 1550~nm laser source, and used a mode cleaner to improve spatial alignment. The multiplexed reference and signal were then sent across a 2.4~km retroreflected turbulent atmospheric link at the University of Western Australia, before measuring their relative electric fields at the receiver (see Fig.~\ref{fig:exp_model}). Three experimental campaigns were conducted in July 2025, a morning campaign ($\sim$9:00am), a noon campaign ($\sim$12:00pm), and an evening campaign ($\sim$5:00pm), where we considered seven HG modes ${[\mathrm{HG}_{00}, \mathrm{HG}_{01}, \mathrm{HG}_{10}, \mathrm{HG}_{02}, \mathrm{HG}_{11}, \mathrm{HG}_{20}, \mathrm{HG}_{12}]}$. Using the procedure in~\cite{Dix-Matthews2023}, we estimate the refractive index structure parameter, $C^2_n$, to be $2.7 \times10^{-15}$~m$^{-2/3}$, $2.57 \times 10^{-15}$~m$^{-2/3}$ and $2.68 \times 10^{-15}$~m$^{-2/3}$ for the morning, noon and evening campaigns, respectively. In the MPLC, a series of phase plates transformed the HG modes into $N$ spatially separated Gaussian beams in the HG$_{00}$ mode, which were coupled into separate single-mode fibers (SMFs). The photodetectors shown in Fig.~\ref{fig:exp_model} measured the voltage $V_{mn}$ of each SMF, which had a logarithmic response to the power in each mode $P_{mn}$, as ${P_{mn} = |c_{mn}|^2 = 50V_{mn} - 114.543}$ (in dBm). Note that the set of $V_{mn}$ values represents the actual measurements in our experiments - all other metrics we refer to in this work are derived from these voltages. 

The reference and signal were frequency shifted by ${\Delta f = 50}$~kHz during transmission. This shift formed a heterodyne beat in each HG mode, resulting in derivations of the reference mode powers $P_{mn,R}$, signal mode powers $P_{mn,S}$, and the relative mode phase errors (MPEs) $\Delta\phi_{mn}$, via Hilbert transformation~\cite{Brandwood2012}. The relative MPEs represent the differential phase errors in each HG mode, ${\Delta\phi_{mn} = \Delta\phi_{mn,R} - \Delta\phi_{mn,S}}$, \textcolor{black}{a modal representation of the relative WFEs,} where $\Delta\phi_{mn,R}$ and $\Delta\phi_{mn,S}$ are the arguments of the HG mode complex coefficients, $c_{mn}$, of the reference and signal, respectively. Importantly, we verified that the 50~kHz frequency shift had an inconsequential contribution to the relative MPEs during lab experimentation, when we cycled the signal frequency by $\pm$50~kHz, relative to the reference.





\begin{figure*} [htbp]
    \begin{centering}
    \includegraphics[scale=0.75]{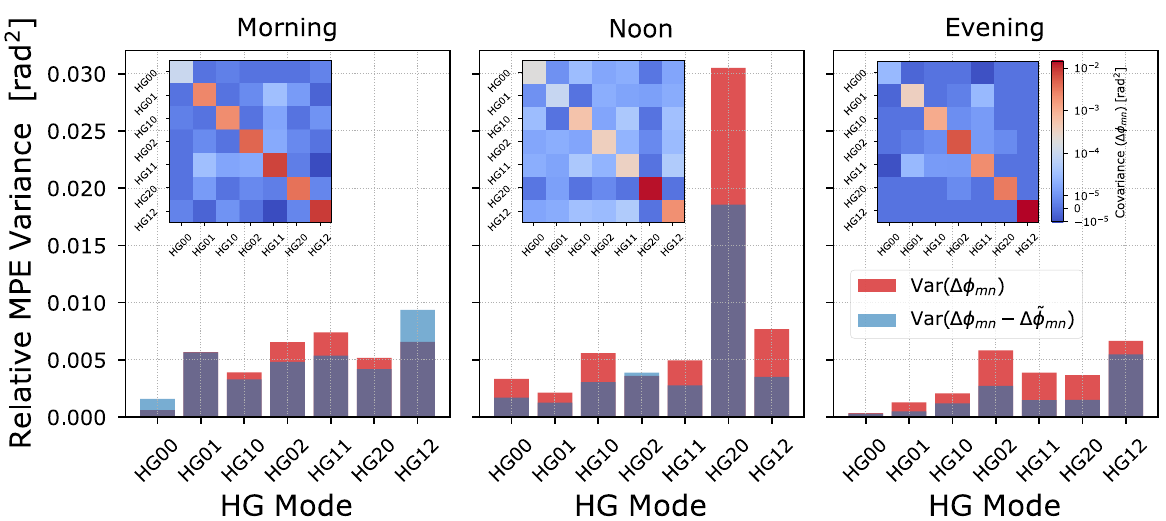} 
    \caption{Derived variances ${\mathrm{Var}(\Delta\phi_{mn})}$ (red bars) versus correction variances ${\mathrm{Var}(\Delta\phi_{mn} - \Delta\tilde{\phi}_{mn})}$ (blue bars) for all campaigns, where purple shows where the derived variances and correction variances overlap. Covariances of relative mode phase errors are shown in the insets. $C^2_n$ was estimated to be $[2.70, 2.57, 2.68] \times10^{-15}$~m$^{-2/3}$ for the morning, noon and evening campaigns, respectively.}\label{fig:dphi_est_var}
    \end{centering}
\end{figure*}

The relative MPE variances derived from our measurements, $\mathrm{Var}(\Delta\phi_{mn})$, are shown as the red bars in Fig.~\ref{fig:dphi_est_var} for all campaigns. The variance was lowest for the HG$_{00}$ mode in the morning and evening campaigns, whilst it was lowest for the HG$_{01}$ mode in the noon campaign. We note for the noon campaign a spike for the HG$_{20}$ mode, which we believe can be attributed to the low signal-mode power observed in the HG$_{20}$ mode for that campaign \textcolor{black}{(an average of $\sim0.3\%$ of total power at the receiver)}. As can be seen, the derived variances were different for each campaign, indicating that relative MPEs will vary not only between experimental setups, but also between campaigns when using the same setup. Focusing on the two-dimensional insets in Fig.~\ref{fig:dphi_est_var}, the off-diagonal terms show the covariances of the modes. It can be immediately seen that there was a weak relationship between the variation of relative MPEs in one mode and the variation of relative MPEs in another. 

We considered whether the statistical distributions of the relative MPEs were different between modes, indicating that the optical hardware and/or atmosphere evolve the transmitted reference and signal HG$_{00}$ mode into higher-order modes in a unique manner. To this end, we applied the Friedman test, a non-parametric test for differences between groups (here, the relative MPEs in each HG mode)~\cite{Macfarland2016}. The null hypothesis that the relative MPE determinations in each mode were of the same statistical distribution was rejected for all campaigns. 
That is, our results indicate that the relative MPEs in the higher-order HG modes were caused by a variational transformation from the transmitted HG$_{00}$ reference and signal modes.



The experimental results shown in Fig.~\ref{fig:dphi_est_var} clearly demonstrate that relative WFEs \textit{can exist} between the references and signals. As such, we developed wavefront correction algorithms based on encoder transformer neural networks (TNNs). The TNNs were designed to take the reference mode powers as input, then output an estimation of the relative MPEs --- based on the concept of phase retrieval (e.g.~\cite{Dong2023}). The TNN architectures were developed using the simulation-based wavefront correction algorithms in~\cite{Long2026_wfe}, \textcolor{black}{which were tested for various channel conditions and WFE types}. However, in this work, we incorporated an additional temporal dimension for the input reference mode powers, such that the previous $M_t=100$ time-steps (\textcolor{black}{after testing various $M_t$ values}) were used to estimate the relative MPEs for the current time-step. The relative MPEs could then be used to theoretically reconstruct an RLO wavefront correction. \textcolor{black}{In practice, the TNN should be trained to account for a breadth of channel and hardware conditions, where retraining would be required for novel conditions.}

During training of the TNNs, known reference mode powers were mapped to known relative MPEs (using $\sim$85,000 determinations). We used a sample rate of~1~kHz, which we assumed was sufficient to capture relative MPE fluctuations for each atmospheric coherence time ($\tau_0 \approx 4$~ms). The TNN weights were all optimized by minimizing a mean squared error loss function, using the Adam optimizer. Considering $N$ modes, we trained the TNNs using a batch size of~16, for 100~epochs, where the batch size represents the number of samples (here, $M_t \times N$) that were mapped from the input to output before the internal weights were updated, while the epochs represent how many times the entire training dataset (the $\sim$85,000 training determinations) were passed through the TNN during optimization. All weights were fixed after training. During operation, the TNNs took known reference mode powers as input, then output an estimate of the relative MPEs (where we tested $\sim$15,000 determinations). 

We quantified the TNN corrections by comparing the derived variances (red bars) with the correction variances ${\mathrm{Var}(\Delta\phi_{mn} - \Delta\tilde{\phi}_{mn})}$ (blue bars) in Fig.~\ref{fig:dphi_est_var} for all campaigns, where purple shows where the derived variances and correction variances overlap. Overall, it can be seen that the correction variances were generally lower than the derived variances.



\begin{figure} [t]
    \begin{centering}
    \includegraphics[scale=0.75]{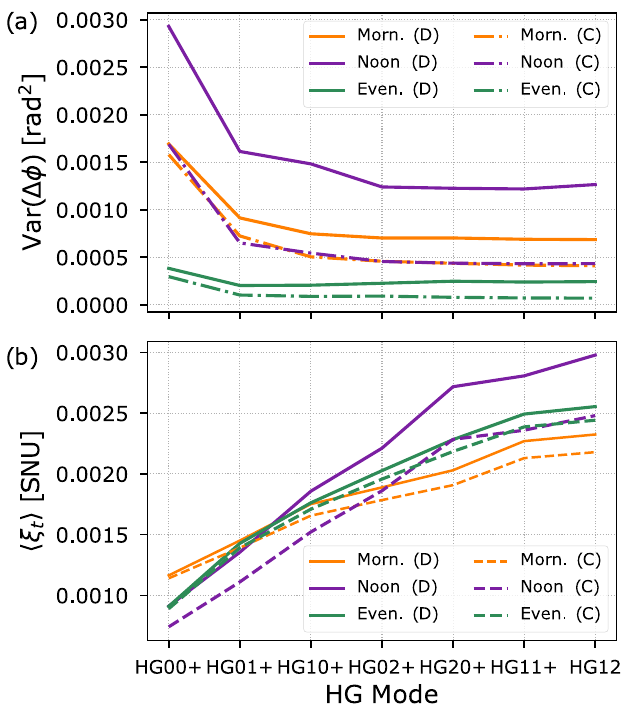}
    \caption{Top-(a) Total relative phase error variances $\mathrm{Var}(\Delta\phi)$ and bottom-(b) average effective excess noise $\xi_t$ for a trusted detector, for accumulating HG modes, for all campaigns, where D represents derived (plotted as solid lines) and C represents corrected (plotted as dot-dashed lines).}\label{fig:xi_eff}
    \end{centering}
\end{figure}

We then extrapolated our experimental results to determine the total relative phase error --- as related to optical communications based on homodyne/heterodyne measurements. Assuming coherent measurement of information-encoded signals with the references, the total relative phase errors, $\Delta\phi$, in the quadrature phase space, were calculated as ${\Delta\phi = \mathrm{arg}(\sum^N_{mn} \sqrt{P_{mn,S} P_{mn,R}} \ e^{i\Delta\phi_{mn}})}$, (see~\cite{Robert2016, Cho2022}). Given the finite number of HG modes, we considered the effect of $N$ on the calculated total relative phase error variance, $\mathrm{Var}(\Delta\phi)$, in Fig.~\ref{fig:xi_eff}(a) for all campaigns, by summing an increasing number of modes (in the order given on the x-axis). Beyond the initial decrease in $\mathrm{Var}(\Delta\phi)$ as $N$ increased, $\mathrm{Var}(\Delta\phi)$ appeared to asymptote. As such, given our experimental results, we assume that the asymptoted $\mathrm{Var}(\Delta\phi)$ were representative of the true $\mathrm{Var}(\Delta\phi)$ (given an infinite number of modes).
When considering the wavefront corrections in Fig.~\ref{fig:xi_eff}(a) for $N=7$ modes, the total $\mathrm{Var}(\Delta\phi)$ decreased by $0.0003$~rad$^2$ for the morning campaign, $0.0008$~rad$^2$ for the noon campaign, and $0.0002$~rad$^2$ for the evening campaign. As such, given $N=7$ modes, we can expect up to a 2/3 decrease in total $\mathrm{Var}(\Delta\phi)$ after applying the wavefront correction, as achieved in the noon campaign.

A target application of this work is Gaussian-modulated coherent state CV-QKD across turbulent atmospheres, potentially to and from satellites~\cite{Hosseinidehaj2019}. Although the signal powers we transmitted across our 2.4~km link were too high to be considered legitimate quantum signals --- we now discuss what impact our results would have on a true CV-QKD experiment if we were to simply assume that in such an experiment, the same relative WFE results and corrections were found. 

CV-QKD across turbulent atmospheric channels often involves encoding information on the electric field quadratures of very weak optical pulses (the quantum signals), polarization-multiplexing them with references during transmission, then using the reference information to modulate an RLO at the receiver, for coherent detection of the quantum signals~\cite{Long2023_survey}. To make progress, we focus on the impact of relative MPEs on the effective excess noise in CV-QKD (all values shown henceforth in shot noise units). There are three main contributions to effective excess noise: channel noise $\xi_{ch}$, detector noise $\xi_{d}$, and channel transmissivity $T$. The effective excess noise can be calculated for a trusted detector as ${\xi_{t} = \xi_{ch} T}$ and an untrusted detector as ${\xi_{ut} = \xi_{ch} T + \xi_{d}}$. A trusted detector assumes that any eavesdropper (Eve) does not have access to the detector - and we focus on that scenario here.

The total relative phase errors, calculated for our experimental relative MPEs, would have a significant effect on phase noise $\xi_{\phi}$ in CV-QKD~\cite{Marie2017}. We calculated phase noise as a function of the total $\mathrm{Var}(\Delta\phi)$ as ${\xi_{\phi} = 2V_{mod}(1-e^{-Var(\Delta\phi)/2})}$~\cite{Marie2017}, for the signal modulation variance $V_{mod}$, \textcolor{black}{resulting in $\xi_{\phi}=0.0034$ for the morning campaign, $\xi_{\phi}=0.0063$ for the noon campaign, and $\xi_{\phi}=0.0012$ for the evening campaign, for $N=7$ modes before the wavefront correction was applied. We then calculated the channel noise as a summation of phase noise and the following estimated values over an Earth-satellite channel (adopted from~\cite{Kish2020})}: background noise $\xi_{back}=0.0002$, modulation noise $\xi_{mod}=0.0005$, time-of-arrival fluctuations $\xi_{ta}=0.006$, the relative intensity noise (RIN) of the signal $\xi_{RIN,S}=0.0001$, RIN of the reference $\xi_{RIN,R}=0.01$, and RIN of the RLO $\xi_{RIN,LO}=0.0018$, for a modulation variance of $V_{mod}=5$. \textcolor{black}{Note that of these noise terms, time-of-arrival fluctuations would be one mechanism that impacts the reference and signals differently. We do not claim that this is the source of the relative WFEs observed --- we simply point out that the levels we see in the experiment are consistent with those anticipated for time-of-arrival effects~\cite{Wang2018}. } Note also that we calculated the transmissivity for increasing $N$ as ${T = (\sum^N_{mn} P_{mn,R})/P_{T,R}}$, where we focused on the reference mode powers and the reference transmit powers $P_{T,R}$. The average transmissivity increased from ${T \approx 0.043}$ for ${P_{00,R}/P_{T,R}}$ (for~$N=1$) to ${T \approx 0.115}$ for ${N=7}$ modes.

We consider the effect of $N$ on the average effective excess noise $\langle \xi_t \rangle$ in Fig.~\ref{fig:xi_eff}(b) for all campaigns, by summing an increasing number of modes (in the order given on the x-axis). Here, two variables are changing as $N$ increases: the transmissivity and the total $\mathrm{Var}(\Delta\phi)$. It can be seen that the average effective excess noise \textit{increased} as $N$ increased, even though the total $\mathrm{Var}(\Delta\phi)$ decreased as $N$ increased. Therefore, the average effective excess noise was dominated by the increase in transmissivity as $N$ increased. Without considering the total impact of transmissivity on secure key rates in CV-QKD, this would indicate that shaping the RLO with only the HG$_{00}$ mode would result in the best performance, where higher-order orthogonal HG modes should be discarded. 

Implementation of our wavefront correction algorithm, which leads to a reduction in effective excess noise, helps to compensate for the additional effective excess noise caused by the increased transmissivity. For $N=7$ modes, the effective excess noise was reduced by $0.0002$ for the morning campaign, $0.0005$ for the noon campaign, and $0.0001$ for the evening campaign, representing up to a $17\%$ decrease in the effective excess noise (for the noon campaign). Therefore, implementation of our wavefront correction algorithms could help mitigate the destructive impact of relative WFEs on effective excess noise, which can have a detrimental effect on secure key rates, in the context of CV-QKD across atmospheric turbulent channels. For example, previous work has shown nearly an order of magnitude increase in secure key rates for excess noise reductions of 15$\%$~\cite{Kleis2019}.

In summary: We have presented experimental evidence of relative WFEs between multiplexed optical references and weaker signals, after transmission across a 2.4~km turbulent FSO link. We developed ML-based wavefront correction algorithms to correct for the observed relative WFEs, reducing the \textit{total} $\mathrm{Var}(\Delta\phi)$ by up to a factor of 2/3. The impact of such a phase error reduction in the context of CV-QKD was presented, showing that an order-of-magnitude increase in QKD secure key rates is potentially achievable using these corrections. \textcolor{black}{Our wavefront correction scheme enables the practical implementation of RLO-based CV-QKD without introducing additional security vulnerabilities beyond those inherent to the underlying CV-QKD protocol. In particular, attacks targeting the reference signal do not introduce any new security issues.}
While we did not identify the source(s) of the relative WFEs and whether the channel itself is contributing to them, \textcolor{black}{it is assumed that hardware imperfections are their primary source, given low birefringence of FSO channels.} However, the identification of the WFE source(s) is not a necessary component of our correction scheme.

\begin{backmatter}

\bmsection{Acknowledgment}
The Commonwealth of Australia (represented by the Department of Defence) supports this research through a Defence Science Partnerships agreement. A.F. and J.W. are supported by Australian Government Research Training Program Scholarships.

\bmsection{Data Availability Statement}
Data underlying the results presented in this paper are not publicly available at this time but may be obtained from the authors upon reasonable request.

\bmsection{Disclosures}
The authors declare no conflicts of interest.


\end{backmatter}


\bibliography{bibliography}

\newpage

\bibliographyfullrefs{bibliography}

\end{document}